\documentclass[a4paper]{article}

\pdfoutput=1 

\usepackage[T1]{fontenc}
\usepackage[utf8]{inputenc}
\usepackage[english]{babel}
\usepackage{graphics}
\usepackage{epsfig}
\usepackage{amsmath}
\usepackage{amssymb}
\usepackage{amsthm}
\usepackage[breaklinks=false,colorlinks]{hyperref}
\usepackage{float}
\usepackage{pifont}
\usepackage[table]{xcolor}
\usepackage{lmodern}
\usepackage{xspace}
\usepackage{bbold}
\usepackage[capitalize,nameinlink]{cleveref}
\usepackage{textcomp}
\usepackage{enumerate}
\usepackage{footmisc}

\usepackage{draftwatermark} 
\SetWatermarkText{Preprint} 
\SetWatermarkScale{1}       

\definecolor{mygray}{rgb}{0.9,0.9,0.9}

\newcommand{\Program}[1]{\texttt{#1}} 
\newcommand{\Code}[1]{\texttt{#1}} 

\title{\vspace{-3em}Sustainable Research Software Hand-Over}

\author{%
J. Fehr\thanks{Institute of Engineering and Computational Mechanics, University of Stuttgart, Pfaffenwaldring~9, 70569 Stuttgart, Germany
\newline ORCID:\ 0000-0003-2850-1440, \url{joerg.fehr@itm.uni-stuttgart.de}}
~~
C. Himpe\thanks{Computational Methods in Systems and Control Theory, Max Planck Institute for Dynamics of Complex Technical Systems, Sandtorstr.~1, 39106 Magdeburg, Germany
\newline ORCID:\ 0000-0003-2194-6754, \url{himpe@mpi-magdeburg.mpg.de}}
~~
S. Rave\thanks{Institute for Computational and Applied Mathematics, University of M\"unster, Einsteinstrasse~62, 48149 M\"unster, Germany
\newline ORCID:\ 0000-0003-0439-7212, \url{stephan.rave@uni-muenster.de}}
~~
J. Saak\thanks{Computational Methods in Systems and Control Theory, Max Planck Institute for Dynamics of Complex Technical Systems, Sandtorstr.~1, 39106 Magdeburg, Germany
\newline ORCID:\ 0000-0001-5567-9637, \url{saak@mpi-magdeburg.mpg.de}}
}

\date{}

\begin{document}

\maketitle

\section*{Abstract}
Scientific software projects evolve rapidly in their initial development phase,
yet at the end of a funding period, the completion of a research project,
thesis, or publication, further engagement in the project may slow down or cease
completely.  To retain the invested effort for the sciences, this software
needs to be preserved or handed over to a succeeding developer or team, such as
the next generation of (PhD) students.

Comparable guides provide top-down recommendations for project leads.
This paper intends to be a bottom-up approach for sustainable hand-over
processes from a developer's perspective.  An important characteristic in this
regard is the project's size, by which this guideline is structured.
Furthermore, checklists are provided, which can serve as a practical guide for
implementing the proposed measures.

\section{Introduction}
Research software, software artifacts as research products, or computer-based
experiments are drivers of modern science.  Yet, while computerization has
massively accelerated science, the intangible and volatile nature of software
has also inhibited scientific progress: Once-developed-software is often not
usable in subsequent development of algorithms, for example, due to technical
incompatibilities, insufficient documentation, or plain unavailability.  Even
though advances in supplying source codes together with published results are
achieved~\cite{Nat18}, the reusability of such scientific codes remains
unsatisfactory~\cite{JohH18}, and of limited reach when tied to a publication.
So, instead of building on top of ``shoulders of giants'', the ``wheel is
reinvented'' regularly in many branches of sciences and not least in
computational mathematics.  A frequently occurring symptom of this deficiency is
the inadequate treatment of software developed for, or over the course of a PhD
thesis, which may be disregarded either by the original or subsequent developing
PhD candidate.

As scientists, scientific organizations, and funding agencies are becoming more
aware of these issues, guidelines and best practices for good scientific
software conduct are in demand.  Examples for such academically driven
efforts are the guides published by the alliance of German research
associations~\cite{KatF18}, the DFG (Deutsche Forschungsgemeinschaft) ``guidelines for safeguarding good
scientific practice''~\cite{dfg19}, the DLR (Deutsche Zentrum f\"ur Luft- und
Raumfahrt) guideline~\cite{SchMH18}, or the software sustainability institute
guideline~\cite{Jac18}.
These guides present top-down approaches aimed at principal investigators,
decision-makers and coordinators.  Our contribution, on the other hand, intends to
be a bottom-up approach presenting requirements and recommendations for academic
software developers, such as undergraduate students, PhD students, postdoctoral
researchers, or research software engineers.  Furthermore, instead of focusing
on the development process of scientific software, as
in~\cite{Hon14,Hon16,Irv15,FehHHetal16} and references therein, we focus on
the continuation of a project, when the developer (or a maintainer) leaves,
e.g.~after completing their PhD project.

We note that industry has already adapted robust collaborative software
development practices, see for example~\cite{Hen17}.  Yet, given
that developers of scientific codes may have no formal training in software
engineering, and scientific software development processes can differ,
in academia only certain ideas can be transferred to support researchers or departments.

While the issues addressed in this work apply to all branches of science, we
emphasize that mathematical software projects hold particular responsibilities.
An example are the numerical libraries \Program{BLAS}~\cite{LawHKK79} and \Program{LAPACK}~\cite{Lapack},
which constitute the basis for numerical computations in many sciences.
Hence, authors of this foundational layer in scientific software stacks need to
take into account the continued use and possibly further development outside the field of mathematics.
Best practices for mathematical software~\cite{Ric71,CroDM79} and numerical
software~\cite{Lau85,Boi97} are long known (yet still not established), and properties
such as reliability, robustness or transportability~\cite{Cod82}, the numerical
experiment attributes replicability, reproducibility and
reusability~\cite{FehHHetal16}, code as a form of scientific
notation~\cite{Tim16,Hin17}, as well as basic guidelines for research
software~\cite{QueSMetal17} have been discussed in the literature, yet, sustainable
hand-over strategies for (mathematical) research software projects have not been
documented to the best knowledge of the authors.

\begin{figure}\centering
 \includegraphics[width=\textwidth]{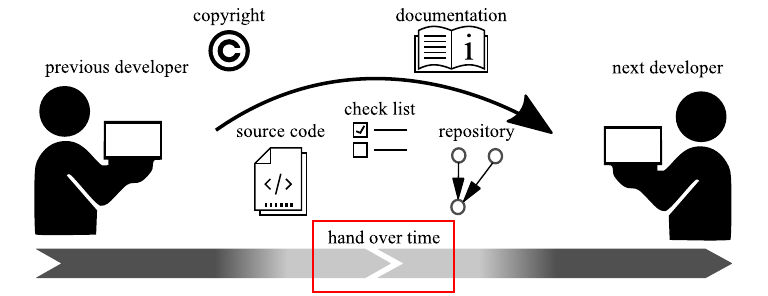}
 \caption{Project hand-over illustrative summary.}\label{fig:fig1}
\end{figure}

The core of this work aims at the hand-over of general scientific software
projects, illustrated in \cref{fig:fig1},
which is discussed in detail in the following sections.
We consider two classes of research software projects:
First, small projects, see \cref{subsec:smallproject},
which are implemented by a single developer, for example over the course of a PhD program or a funding period;
Second, large projects, see \cref{subsec:largeproject},
which have multiple developers.
Since these two project categories serve different purposes, the proposed
requirements and recommendations differ.  Minimal requirements, as well as
optional recommendations, are given for both project categories.  Finally in
\cref{sec:con}, a brief conclusion is given alongside two checklists, which summarizes the proposed
measures for a practical hand-over process,
followed by a brief comment on minimal documentation of numerical software in
the \nameref{sec:app}.

\section{Project Hand-Over}
In the following, we lay out minimal and optional measures for a sustainable
project hand-over distinguished by the size of the project.  From our
experience, we recommend the distinction of software projects into the two
categories ``small'' and ``large''.  A more fine grained categorization is
surmisable too, see e.g.~\cite{INRIA}, still, we think that two categories are
sufficient in covering the essential aspects of sustainable software hand-over, with
the rationale that more straightforward guidelines may have a higher chance of general
acceptance compared to more complicated rule sets.

As a general remark: When a project is handed over, a time period from before the
previous developer leaves, till after the next developer enters the project is
considered the hand-over time, which should be allocated in a manner to suitably prepare the
hand-over, and allow for a training phase.
To this end, it can be worth the extra cost of having the previous and next developer(s) 
overlap for some time, depending on the project size and complexity.
We also note that if a project is not continued in direct succession, it can be
conserved; see for example~\cite{swh}, for information on archiving.

\subsection{Small Project}\label{subsec:smallproject}
We consider a \textbf{small project} to be code developed and maintained by a
single author, which means, for example, a project written from scratch, or a
fork of an existing project that throughout the development is not merged
back into the parent project. This is often the case for tools developed as
part of a publication, thesis or with a tight focus.  Such projects have their
developer as the sole user, or at least a limited user base.

Following, we will lay out minimal requirements, which ensure the project's
sustainability, as well as optional recommendations that facilitate long-term
usability, such as, when a new student takes over, after a previous student
finishes their work, or if an abandoned project is revived.

\pagebreak

\subsubsection{Minimal Requirements}

\paragraph{Code availability}\label{sec:sm1}
The most important requirement for continuation or at least conservation is the
availability of the project contents --- utilized specific hardware components may need to be kept available physically, if no virtualization is possible --- including the source code, configuration and data files.
Therefore, the project location should be discoverable, i.e.:\ not solely on the
developer's personal computer hard-drive, but rather in a central repository of the associated
institute at a known and accessible storage location.

\paragraph{Code ownership}\label{sec:sm2}
If the code is available, the next important question is: Who owns the code?
Potential owners could be the associated institute or university, the superior
or supervisor of the developer, or the original developer themselves.
Additionally, if there is third-party funding involved, the funding entity may
have regulations about the funded project's ownership.  Besides ownership,
third-party rights need to be considered, originating from prior developers,
third-party projects, or parts thereof included in the project.  These ownership
question can be resolved by documentation of stakeholders alongside the code and
with a license statement, which can be as easy as the project's developer
self-licensing their work or following the respective guidelines applicable
to them.  For further information on software licensing see~\cite{Sto09}.

\paragraph{Execution environment}\label{sec:sm3}
Given all legal prerequisites are resolved, a minimal description of the
required runtime environment, such as operating system, dependencies, and
compiler or interpreter is needed, together with a short description on how to
compile, if necessary, and run the project.  A tested upon operating system
needs to be stated (with compute architecture and endianess if
applicable). 
We also recommend listing all depending software libraries, tools or toolboxes,
which are not part of the default installation of the compatible operating systems.  Furthermore, all
components of the required software stack need to be given with a version
number.  We caution that even in case of high-level cross-platform runtime
environments, certain behavior may depend voluntarily, accidentally, or due to
restrictions, on the underlying operating system (for a minimal report, in this case, see the \nameref{sec:app}).
In view of increasingly complex scientific computing software stacks (\cref{fig:tower}),
providing a reproducible execution environment (see below) is highly recommended.

\begin{figure}\centering
  \includegraphics[height=.35\textheight]{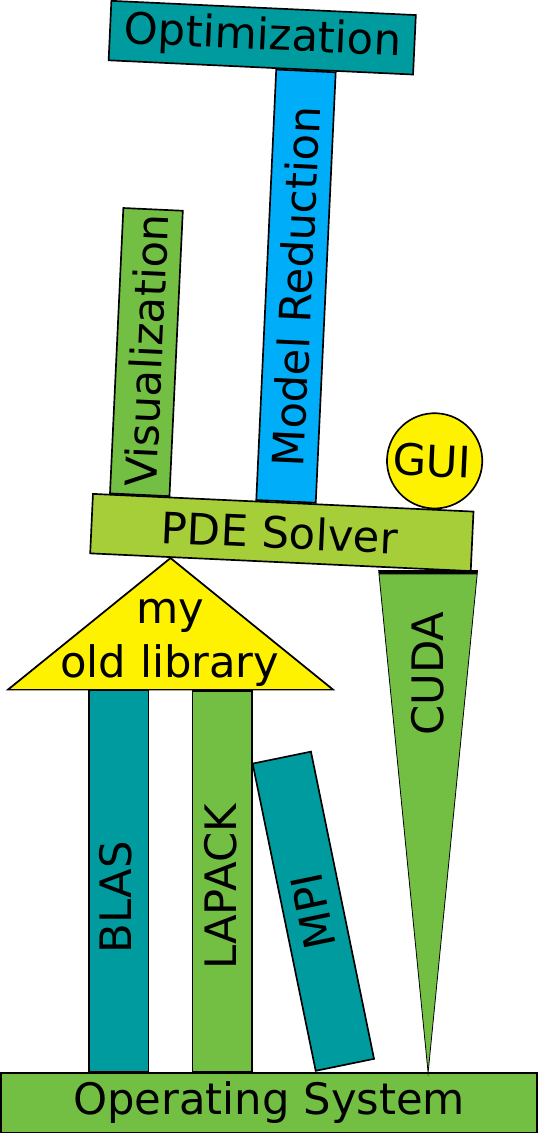}
  \caption{Software stack dependencies: ``Tower of Doom''.}\label{fig:tower}
\end{figure}

\paragraph{Working example}\label{sec:sm4}
An essential requirement for a small project hand-over, is sample code
(In~\cite{FehHHetal16} such a file is suggested to be named \Code{RUNME}.),
which can run and demonstrate the core feature(s) of the project.  Such an  
example is essential, to test if the code is executable and also serves as a
starting point to understand the structure of the code, since the workflow can
be traced for a known working example, e.g.\ by a debugging program.  Moreover,
the results can be used to verify that future changes do not (unintentionally)
affect computational results. To these ends, the execution of such an example code should
sufficiently cover the complete functionality of the software project.

\pagebreak

\paragraph{Minimal documentation}\label{sec:sm5}
Typically the information of the previous requirements is gathered in a
\Code{README} file (\Code{README} is a widely used file name for a
  plain text file, holding a minimal documentation;
  see:~\cite{FehHHetal16}). Further information that should be included in the
\Code{README} is:


\begin{itemize}
 \item Is the code functioning, and if, on what hardware (see \nameref{sec:app})?
 \item Is the available project state current (latest use in a thesis or publication)?
 \item New algorithms from which publications are implemented by this project?
 \item Existing algorithms from which publications are utilized by this project?
 \item What publications use this project?
 \item What are the known limitations or issues?
\end{itemize}
Referencing all associated publications helps to put a small research software
project in the appropriate scientific context, and has also educational function
for the subsequent developer(s). 

\subsubsection{Optional Recommendation}

\paragraph{Public release}\label{sec:so1}
As the availability of the project is crucial, for the documentation of the
scientific findings, the best measure is a public
release under an, ideally, open license on a stable service~\cite{CosZ17}.  If
legal or other reasons prevent such line of action, the reasons should be stated
near the top of the aforementioned \Code{README} file, so this important
information is not lost in transition.

\paragraph{Version control}\label{sec:so2}
We strongly recommend to use a version control software to track the changes during the
development of the project in a repository. Besides documenting the history of
a project, modern version control systems allow to tag (mark) states of the
repository.  This is useful for associating experiments, for example in
publications, during the development process.  Hence, all experiments can refer
to a specific revision of the source code, in order to ensure replicability and
reproducibility, in particular for future developers.  At the very least a
version control repository serves as a (very sophisticated) back up method. An
introduction to generic version control workflows can be found in~\cite{Wes15}.

\paragraph{Basic code cleanup}\label{sec:so3}
Furthermore, some software development anti-patterns~\cite{BroMMetal98} are more
common (in our experience) in small projects, and impede project continuation by
another than the original developer.  First, undocumented constants used in the
source code hinder the interpretation in the absence of the original developer.
Second, comments containing code, so called dead code, introduce the uncertainty
which code has been used for what experiments, and if the commented out code is
still needed or not.  Third, the use of hard-coded file paths may prevent the
project from functioning in a different environment, such as another developer's
computer.  All these issues can, if not fixable, be easily resolved by a few
additional source code comments.

\paragraph{Reproducible execution environment}\label{sec:so4}
In addition to the minimally required documentation, we recommend to report if
the project was tested in other compute environments than the developer's.  To
ensure long term compatibility and conservation, it is relevant if the project
can run on a simulated computer, i.e.\ a virtual machine.  This allows
conserving an image file, treated as a hard drive by such a virtual machine, containing
the complete software stack (including the operating system).  Thus, the image
file completely defines the software aspect of the compute environment, and the
virtual machine software presents an abstraction from the hardware.

As an alternative to a virtual machine image, a step-by-step guide can be
included, which explains the preparation, i.e.\ correct sequence of installation
of dependencies, starting from the base installation of a compatible operating
system.  Such a guide can be easily distributed with the software, whereas, due
to their size, virtual machine images often need to be archived separately.
Moreover, the guide can serve as a starting point for installing the software in
other execution environments.

\paragraph{Integration into larger project}\label{sec:so5}
A possible path for small projects is the inclusion into existing larger projects,
which, for example, provide a collection of topically related functionality, like a community library.
Such a large project mitigates some of the aforementioned problems due to development guidelines.
To be included into the code base of such a super-project,			
it is essential for the small project to be modular
and compatible with the including project's principle design, interfaces, style and contribution guidelines,
as well as possibly build and test systems.
Furthermore, planned or unsuccessful directions of development should be
included into the documentation to support the future (third-party) development
of the incorporated small project.

Practically, there are three paths to include a smaller project into an overarching project:
First, the continuous development, for example, as a feature of the large project.
This approach naturally requires adherence to project guidelines and often entails slower progress due to this overhead.
Second, after completion, requesting inclusion of the finished ``small project'';
while quick progress can be made this way during development,
integration may be hard due to independent design and build systems.
Third, a fork of the super-project with subsequent independent development,
and a final merge, which may also require some adaptions,
likely allows efficient development without giving up the frame of the super-project.

Alternatively, if direct integration into the large project's code base is impractical,
preserving the small project as a module or callable library,
together with integration of a binding interface into the large project, can be an option.
To track the large project's dependence on a specific version of the small project,
version control system features such as submodules or subrepositories can be used.

\subsection{Large Project}\label{subsec:largeproject}

We define a {\bfseries large project} as a software package that is 
developed by multiple authors, possibly located at different institutions. An
example setting is a project consortium developing a joint tool driven by their
research that also should be made available, e.g.\ to their peers. While the
developing researchers may be a significant subgroup of the software's users, in
this case the community can be far larger and the users might even be unrelated
to this community. 

\begin{figure}\centering
	\includegraphics[width=.95\textwidth]{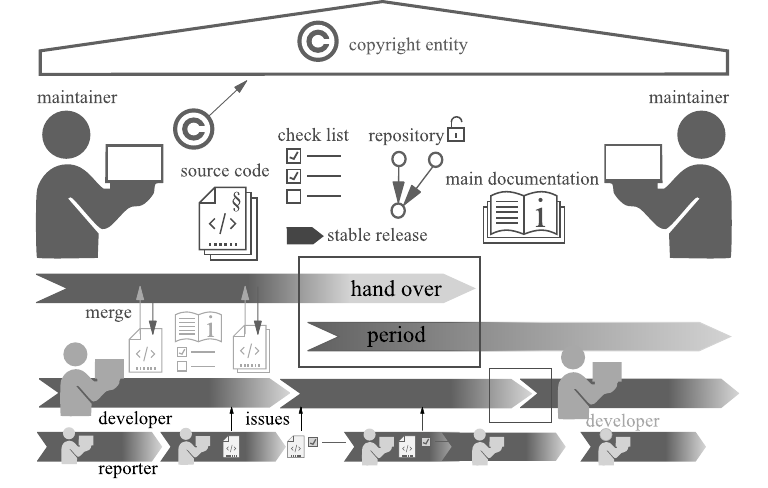}
	\caption{Project hand-over illustrative summary for a larger project.}\label{fig:fig3}
\end{figure}

In our experience it is advisable that large projects have a hierarchy of
contributors, see \cref{fig:fig3}, which follows de-facto standards.
Unprivileged users serve as
{\bfseries reporter}s, who file feature requests or bug reports (which can jointly
be called issues).  Contributors that work on closing bugs or
contributing features are called {\bfseries developer}s.  They have limited, or
no write access to the main development line of the software.  The {\bfseries
  maintainers} have extended permissions on the
repository and oversee the progress of the software project.  They also merge
the contributions of the developers into the main development line.  
While reporters and developers may change frequently, maintainers ensure consistency
of the development, at most superseded by a rights holding entity, depicted in \cref{fig:fig3} as a roof of the project.

In the following sections, we propose hand-over guidelines for large
projects, subdivided into bare minimum requirements and optional, but desired,
recommendations.  While for developers the guidelines for small
projects (\cref{subsec:smallproject}) apply to their branches (a \textbf{branch}
is a copy of the development resources under version control which can evolve in
separate, but is still part of the overall source code repository.), the
presentation, here, focuses on maintainers.

\subsubsection{Minimal Requirements}

\paragraph{Software license}\label{sec:lm1}
The chosen project license is important, even crucial for publicly available
projects.  While for a small project only few entities are eligible to act as
the rights holder, for large projects the situation can be, and often is, more
complex.  This, in turn, leads to additional difficulties that need further
attention: Project funding can end after a certain period, and maintainers
may change their employers or even fields of interest.
Thus, to ensure continued availability of the project,
the developers need to come to a formal agreement, i.e.\ a software license,
under which terms the project should be available.
For an open-source license hierarchy, see~\cite{Whe07}.

\paragraph{Code ownership of contributions}\label{sec:lm2}
Compared to small projects, the question of contributed code's legal ownership is more
relevant for large projects.  In particular, developers need to consider that
a later change of license requires the consent of all copyright holders, which
may have long left academia.
Therefore, if a license change shall remain feasible, all code contributors
could transfer their copyright to a single entity, for example,
a society or association as copyright holder.  It should
also be noted, that there are important differences in copyright laws over the
world and obtaining proper legal advice is desirable.

\paragraph{Access to project resources}\label{sec:lm3}
Similarly important as legal rights are the access permissions in the software
repository and further project resources, such as servers,
websites, domain names or mailing lists.  As a minimal requirement, there should
always be at least two persons with administrator access to all project
resources.  In case of a smaller development team with only one active
maintainer, it is sufficient if these rights are held by a second person who
is associated with the project but is not an active developer (like a research
group leader).  This measure prevents a project depending on the health and
goodwill of a single individual.

\paragraph{Management of development branches}\label{sec:lm5}
Modern version control systems permit ways to continue developing a version of
the software independently from a given state of the main development stream,
e.g., for development of new features.  These are called {\bfseries branch}es,
and it is good practice to use one branch per user, or issue. Each branch has to
be documented with respect to its purpose and status; furthermore, it should be
clear which developers are responsible for the branch.  If the withdrawal of a
developer from the project leads to an unmaintained branch, the branch should
either be merged into the main development branch, a new developer for the
branch should be found, or in case either is not feasible, a detailed
description of the open and completed tasks should be added to the
documentation to allow continuation after a stale phase.

\paragraph{Stable main branch}\label{sec:lm4}
To ensure that a leaving maintainer cannot cause an unknown or unusable state of
the project, it is essential to make sure that the main branch of the software
can be (if applicable compiled and) executed by more than a single person (the
main developer) and runs on all targeted platforms at any time during the
development process.  This also means that the installation is flexible enough to at least
specify user-specific paths during the build process.

\subsubsection{Optional Recommendation}

\paragraph{Division of responsibilities}\label{sec:lo0} 
Beyond a certain project extent, distributing the workload among multiple maintainers may become necessary:
Scientific software projects often comprise segregated functional compartments,
for example: reading, processing, storing, visualizing, or forwarding data.
Depending on their complexity, all these steps may branch into a large variety of available methods,
and thus be too complex to oversee in detail for a single maintainer.
In this case maintainers should be assigned for different parts of the project,
and their responsibilities recorded in the code repository, e.g.~\cite{github_codeowners}.
Whenever a maintainer leaves, their responsibilities need to be handed-over to another maintainer.

\paragraph{Code maintainability}\label{sec:lo1}
All measures that improve the overall quality of the code and its
maintainability are also beneficial in a hand-over process as they facilitate
the familiarization of a new developer with the project.  More importantly,
after the withdrawal of a developer, old code that has been written by this
developer will be much easier to understand if standard software development
best practices are followed. In particular, we mention usage of continuous
integration (CI). In software engineering, continuous integration is the
practice of merging all developers' working copies into the main development
line regularly. This is often followed by a test-phase to ensure that
none of the recent changes break other functionality (see also~\cite{Bec03}). An
optional add-on, which is especially relevant for scientific computing software,
is the more recent technique of continuous benchmarking that
additionally tries to ensure optimal performance of the implementation at all
times. Furthermore, if applicable, we recommend the usage of build systems that
automatically resolve dependencies, especially to other projects, during the
compilation process.

\paragraph{Changelog}\label{sec:lo2}
As soon as a software is developed and used by more than one person, keeping
track of important changes in the software compared to earlier versions becomes
consequential.  While the history of version control systems allows inspecting every
change of the software, this information is usually too fine grained for the ``big picture''.
Therefore, the most relevant changes should be documented in a
\Code{CHANGELOG} file~\cite{kac} or the release notes.
This document not only informs users about new features, the removal of
faulty code or changes in the interfaces, but also helps developers of other
software projects relying on the 
function interfaces, to keep track of changes and necessary updates to their own
projects.  More importantly in the scope of a project hand-over it is helpful for the new
maintainer to comprehend changes and note dependencies as well as compatibilities,
especially if legacy versions of a project need to be maintained, e.g.\ due to 
hardware restrictions, in parallel to the evolution in the main development branch.

\paragraph{Code of conduct}\label{sec:lo3}
A document defining rules for the introduction and retirement of project
maintainers as well as handling project administration questions can have an essential
role in project hand-over.  In particular, when a maintainer no longer actively
works on the project but is hesitant to step down, a code of conduct document
can prevent an entailing gridlock in the project.

\paragraph{Contribution policy}\label{sec:lo4}
Besides the legal status of contributions discussed above,
a contribution policy defines the practical requirements for the contributed code.  Typical
requirements regard the general workflow of the project. For example,
requirements state whether single or multiple pull/merge requests, with what
level of documentation and tests, are expected. 
The code should be mergeable with the main development branch. Also, (passing) tests for all included features
can be expected in the project's favored test suite. The licensing and copyright
of the contributed code as well as the 
form of attribution of the contribution should be clear. 
Oftentimes also restrictions on the code's general layout and naming schemes are
prescribed, in order to improve readability and thus accessibility of the
implemented ideas.
 
 As discussed above, a
case of project hand-over is the inclusion of a smaller into larger project.
Such a policy can simplify this process, in particular, if these requirements
are known during the development of the small project.

\begin{table}
  \small
  \renewcommand{\arraystretch}{1.5}
  \renewcommand{\tabcolsep}{0.25em}
  \begin{tabular}{cl@{\hskip 1em}p{.55\linewidth}}
    \multicolumn{3}{l}{\textbf{Small Software Project Handover}}\\
    \hline\\[-0.5em]
    $\blacksquare$&\multicolumn{2}{p{\linewidth}}{\textbf{Minimal Requirements}} \\

    $\square$&\nameref{sec:sm1} & Where are source code, data and configuration files? \\
    
    $\square$&\nameref{sec:sm2} & Who owns the software and who holds rights? \\
    
    $\square$&\nameref{sec:sm3} & What hardware and software stack is required? \\
    
    $\square$&\nameref{sec:sm4} & How are the features of the code producing what results? \\
    
    $\square$&\nameref{sec:sm5} & What does a new developer need to know at
                                  the least? \\[4em]
    
    $\blacksquare$&\multicolumn{2}{l}{\textbf{Optional Recommendations}} \\
    $\square$&\nameref{sec:so1} & Is a public open-source release possible? \\

    $\square$&\nameref{sec:so2} & Are revisions of the software automatically tracked? Where? \\
    
    $\square$&\nameref{sec:so3} & Are constants, dead code and hard paths removed? \\
    
    $\square$&\nameref{sec:so4} & Is a (virtual) machine back up available? \\
    
    $\square$&\nameref{sec:so5} & Is inclusion into a larger project possible or planned?
                                  
  \end{tabular}

\caption{Checklist for sustainable research software hand-over of \textbf{small} projects.}\label{tab:checklistsmall}
\end{table}

\begin{table}
  \small
  \renewcommand{\arraystretch}{1.5}
  \renewcommand{\tabcolsep}{0.25em}
  \begin{tabular}{cl@{\hskip 1em}p{.53\linewidth}}
    \multicolumn{3}{l}{\textbf{Large Software Project Handover}}\\
    \hline\\[-0.5em]
    $\blacksquare$&\multicolumn{2}{p{\linewidth}}{\textbf{Minimal Requirements}} \\
    
    $\square$&\nameref{sec:lm1} & Has a suitable (and compatible) software license been chosen? \\
              
    $\square$&\nameref{sec:lm2} & Who owns which parts of the code? \\
              
    $\square$&\nameref{sec:lm3} & Are full permissions to all project resources granted to at
                                   least two persons? \\
    $\square$&\nameref{sec:lm5} & Are there unmaintained development branches? \\
    $\square$&\nameref{sec:lm4} & How is stability of the main branch ensured?\\[4em]

    $\blacksquare$&\multicolumn{2}{l}{\textbf{Optional Recommendations}} \\

    $\square$&\nameref{sec:lo0} & Do all parts of the project have a responsible
                                  maintainer? \\

    $\square$&\nameref{sec:lo1} & Is continuous integration / testing / benchmarking utilized? \\
              
    $\square$&\nameref{sec:lo2} & Are the core changes of the releases tracked in a changelog or release notes? \\
              
    $\square$&\nameref{sec:lo3} & What are the central points of the code of conduct and why? \\
              
    $\square$&\nameref{sec:lo4} & How are contribution policies communicated?

  \end{tabular}

\caption{Checklist for sustainable research software hand-over of \textbf{large} projects.}\label{tab:checklistlarge}
\end{table}

\section{Sustainable Hand-Over}\label{sec:con}


In this work we presented measures for the sustainable hand-over of research
software, by differentiating between small and large software projects and
proposing minimal requirements and optional recommendation for both categories.
With this, we aim to spark a discussion in the sciences on sustainability of
research software development and appreciate feedback.  Furthermore, we hope
that this document, and especially the checklists in \cref{tab:checklistsmall}
and \cref{tab:checklistlarge} help software sustainability (maybe even beyond science) or at least serve as
a template prototype.

Alternative strategies to academic development, which can also ensure sustainable development,
such as commercialization, were not discussed, as the requirements for small and large projects alike,
first and foremost involve legal issues.
Nonetheless, also in case of academic research software hand-overs,
it is always advisable to consult the involved entity's legal department(s),  
due to the complex situation with copyright, licensing and ownership.


\section*{Appendix}\label{sec:app}
Due to background of the authors, we give some specific documentation hints for
numerical software; this automatically includes code written in the languages
\Program{MATLAB}/\Program{Octave}, \Program{Python}
(\Program{NumPy}/\Program{SciPy}), \Program{R}, and \Program{Julia},  as well as
most research software depending on numerical computations.
The bare minimum information on the computation environment for
these non-compiled numerical software is given by:
\begin{itemize}
 \item Runtime interpreter name and version.
 \item Operating system name, version and architecture / word-width.
 \item Processor name and exact identifier.
 \item Required amount of random access memory.
 \item BLAS library implementation name and version.
 \item LAPACK library implementation name and version.
\end{itemize}
Obviously, in other sciences additional minimal information may be
necessary. For example in lab-sciences hardware and protocols for access to lab
equipment providing the processed data would be essential information.

\section*{Acknowledgments}
Supported by the German Federal Ministry for Economic Affairs and Energy,
in the joint project: 
``\textbf{MathEnergy} --- Mathematical Key Technologies for Evolving Energy Grids'',
sub-project: Model Order Reduction (Grant number: 0324019\textbf{B}).

Funded by the Deutsche Forschungsgemeinschaft (DFG, German Research Foundation) under Germany's Excellence Strategy EXC 2044--390685587,
Mathematics M\"unster: Dynamics--Geometry--Structure.

Supported by the German Federal Ministry of Education and Research (BMBF) under contract 05M18PMA.

The authors would like to thank \textsc{Arnim Kargl} for his help in preparing the hand-over illustrations.

\bibliographystyle{plainurl}
\bibliography{rrr2,additional}

\end{document}